\documentclass[letterpaper, 10 pt, conference]{ieeeconf}
\IEEEoverridecommandlockouts
\usepackage{cite}
\newcommand{\req}[1]{\eqref{#1}}
\newcommand{\rfig}[1]{Fig.\,\ref{#1}} 
\newcommand{\rrem}[1]{Remark\,\ref{#1}} 
\newcommand{\rsec}[1]{Section\,\ref{#1}}
\usepackage[linesnumbered, ruled]{algorithm2e}
\usepackage{bm}
\usepackage{color}
\usepackage{subfigure}
\usepackage{comment}
\usepackage{amsmath,amssymb,amsfonts}
\usepackage{algorithmic}
\usepackage{graphicx}
\usepackage{textcomp}
\usepackage{xcolor}
\newcommand{\qedwhite}{\hfill \ensuremath{\Box}}
\newcommand{\rpro}[1]{Problem\,\ref{#1}}
\newcommand{\ralg}[1]{Algorithm\,\ref{#1}}

\newtheorem{problem}{Problem}
\newtheorem{remark}{Remark}
\newtheorem{example}{Example}
\definecolor{airforceblue}{rgb}{0.36, 0.54, 0.66}

\newtheorem{Specification Set}{Candidate Specifications} 
\def\BibTeX{{\rm B\kern-.05em{\sc i\kern-.025em b}\kern-.08em
    T\kern-.1667em\lower.7ex\hbox{E}\kern-.125emX}}
\begin{document}
\setlength\textfloatsep{9pt}
\setlength\intextsep{9pt}
\title{STL2vec: Signal Temporal Logic Embeddings for Control Synthesis With Recurrent Neural Networks
}
\author{Wataru Hashimoto, Kazumune Hashimoto, and Shigemasa Takai
\thanks{Wataru Hashimoto, Kazumune Hashimoto and Shigemasa Takai are with the Graduate School of Engineering, Osaka University, Suita, Japan (e-mail: hashimoto@is.eei.eng.osaka-u.ac.jp, hashimoto@eei.eng.osaka-u.ac.jp, takai@eei.eng.osaka-u.ac.jp)
}}

\maketitle

\begin{abstract}
In this paper, a method for learning a recurrent neural network (RNN) controller that maximizes the robustness of signal temporal logic (STL) specifications is presented. 
In contrast to previous methods, we consider synthesizing the RNN controller for which the user is able to select an STL specification arbitrarily from multiple STL specifications.
To obtain such a controller, we propose a novel notion called \textit{STL2vec}, which represents a vector representation of the STL specifications and exhibits their similarities. 
The construction of the STL2vec is useful since it allows us to enhance the efficiency and performance of the RNN controller. 
We validate our proposed method through the examples of the path planning problem.
\end{abstract}

\begin{keywords}
Signal temporal logic, neural network controller, optimal control. 
\end{keywords}


\section{Introduction}\label{intro}
Recently, formal methods have attracted much attention in control to address complex and temporal objectives such as periodic, sequential, or reactive tasks, going beyond the traditional control objectives like stability and tracking. Temporal logics such as linear temporal logic (LTL)\cite{LTL}, metric temporal logic (MTL)\cite{MTL}, and signal temporal logic (STL)\cite{STL} allow us to write down formal descriptions for these specifications. 

In this paper, we focus on dealing with the STL specifications, which can specify the temporal properties of real-valued signals. 
One of the notable advantages of using the STL specifications in control is that we have access to the quantitative semantics, called \textit{robustness}, which measures how much a signal satisfies the STL specification. The robustness has been used for control purposes in many existing works \cite{MPC1, MPC2,MPC3, smooth1, approx, smooth 2, smooth 3}.
The authors of \cite{MPC1, MPC2,MPC3} proposed a scheme of encoding STL constraints or robustness of the STL specifications by mixed-integer linear constraints, and these are utilized to employ model predictive control (MPC). 
To alleviate the computational burden of mixed-integer programming, the works \cite{smooth1,approx,smooth 2,smooth 3} proposed the concept of \textit{smooth} robustness, allowing us to solve the maximization problem of the robustness based on gradient-based algorithms. 

Aiming to further improve the scalability and meet real-time requirements, neural network (NN) based feedback controller was employed for STL tasks \cite{NN, DQN, demonstration, RNN1, RNN2}.
Although the learning procedure itself may be time-consuming, once this computation can be done offline, applying the resulting controller online becomes much faster than directly solving the optimization problem at each time step.
In \cite{NN}, the parameters of the feed-forward NN were trained to maximize the robustness of the STL specifications.
Reinforcement learning algorithms such as deep Q-learning (DQN) and learning from demonstrations (Lfd) were used for learning the NN controller in \cite{DQN} and \cite{demonstration} respectively.
The works of \cite{RNN1,RNN2} used a recurrent neural network (RNN) \cite{LSTM} instead of a feed-forward NN, which is more suitable for control with STL specifications since the satisfaction of an STL formula depends on a state trajectory (not just the current state).  
In \cite{RNN1,RNN2}, RNN parameters were trained using imitation learning and model-based reinforcement, respectively.

However, the afore-cited previous works can deal with \textit{only one} STL specification (i.e., the NN controller is trained for only one prescribed STL specification), and did not explicitly deal with multiple STL specifications.  
In particular, it may be more desirable and flexible in practice that the user is able to choose an STL specification freely from multiple STL specifications, such as the case where a surveillance task in a certain region (building, park, \textit{etc.}) changes from day to day. 
A naive approach to achieve this would be to learn a {set of} NN controllers independently for all the candidate STL specifications, although this would be \color{blue} expensive in terms of memory consumption and computational resources \color{black} if the number of the STL specifications is large (i.e., the number of the parameters to be learned could increase as the number of the candidate STL specifications increases). 
Hence, we here consider learning a {\textit{single}} NN controller, 
in which not only the states of the system but also a chosen STL specification are regarded as inputs to the NN. We in particular focus on learning an {RNN} controller since as previously mentioned, it is suitable for dealing with history dependent nature of the STL specifications. 
The above formulation leads us to the following question to be answered, which has not been investigated in previous works of literature:

\smallskip 
\textit{How should we encode the STL specifications by vectors, so that they are readable to the RNN?}
\smallskip

Naively, we could assign any unique numbers or one-hot vectors to all the STL specifications and use them as the inputs to the RNN. Although these approaches are easy to implement, information about 
the similarities in the specifications cannot be encoded with these schemes.
Thus, the underlying function to be learned can be unreasonably complex as the number of candidate STL specifications increases, which leads to increased computation time for training as well as a critical failure in control execution. 

Motivated by the above issue, we propose a novel scheme for constructing a vector representation of STL specifications, called \textit{STL2vec}, which captures similarities between the STL specifications. 
\color{blue}
The vectors for the specifications will be trained based on Word2vec \cite{skip-gram,word2vec} which is a technique widely known in natural language processing.
Then, the parameters for the RNN are trained by taking the vectors obtained by STL2vec and state trajectory as the inputs (to the RNN) and defining a loss function by the average of negative robustness scores for all the candidate STL specifications. 

\color{blue}
The main contributions of this work are summarized as follows. 
First, we propose a method for obtaining vector representations of STL specifications that capture their similarities in terms of control policy. Although STL2vec is constructed based on the concept of word2vec, which is a technique that maps a word of the natural language onto the vector space, how to generate a dataset to learn appropriate vector representations for STL specifications in control synthesis is not a trivial problem. 
We address the details on how to construct dataset to train appropriate STL2vec in Section~\ref{proposed}. 
Second, by using the vectors generated by STL2vec, we train the controller that can deal with multiple STL specifications with one RNN model. Naively, if one aims to train the NN controller to deal with multiple specifications by simply applying existing NN-based control synthesis methods such as \cite{NN, DQN, demonstration, RNN1, RNN2}, one needs to ready NN models as many as the number of the STL specifications, which leads to large memory consumption and computational resources (for details, see Section~\ref{remark:memory}). 
Moreover, the proposed method has the potential to accelerate the training as will be seen in the case study of Section~\ref{case study}. 
Although there exist other methods that make the NN controller flexible such as \cite{RNN1} and \cite{CBF}, which can handle (unknown) obstacles by using control barrier function, these methods are restricted to the specification of only collision avoidance and thus are not intended to deal with multiple temporal logic specifications. The proposed method in this paper allows us to overcome such limitations; it allows the user to have a flexibility of selecting an STL specification freely from multiple ones, and this will be achieved by constructing STL2vec. 
\color{black}

\section{Preliminaries}\label{pre}
\subsection{System description and notations}
We consider a nonlinear discrete-time system of the form: 
\begin{align}\label{dynamics}
    x_{t+1} = f(x_t, u_t),\ x_0 \in \mathcal{X}_0,\ u_t\in \mathcal{U},
\end{align}
where $x_t\in \mathbb{R}^n$ and $u_t\in \mathcal{U}$ are the state and the control input at time $t \in \mathbb{Z}_{\geq 0}$, $\mathcal{X}_0 \subset \mathbb{R}^n$ is the set of initial states, $\mathcal{U} \subset \mathbb{R}^m$ is the set of control inputs, and $f: \mathbb{R}^n \times \mathcal{U} \rightarrow \mathbb{R}^n$ is a function capturing the dynamics of the system. We assume that the initial state $x_0$ is randomly chosen from $\mathcal{X}_0$ according to the probability distribution $p : \mathcal{X}_0 \rightarrow \mathbb{R}$, and $\mathcal{U}$ is given by $\mathcal{U} = \{u \in \mathbb{R}^m : u_{\min}\leq u \leq u_{\max} \}$ for given $u_{\max}, u_{\min} \in \mathbb{R}^m$ (the inequalities are interpreted element-wise). 
Given $x_0 \in \mathcal{X}_0$ and a sequence of control inputs $u_0, \ldots, u_{T-1}$, we can generate a unique sequence of states according to the dynamics \req{dynamics}, which we call a \textit{trajectory}: $x_{0:T} = x_0, x_1, \ldots, x_T$. 
\color{black}
\subsection{Signal Temporal Logic}
Signal temporal logic (STL) \cite{STL} is a logical formalism that can specify temporal properties of real-valued signals. 
The syntax of the STL formula is recursively defined as follows:
\begin{align}
    \phi ::= &\top \mid \mu \mid \neg \phi \mid \phi_1 \land \phi_2 \mid \phi_1 \lor \phi_2 \mid \bm{F}_{I}\phi \mid \notag \\  
    & \bm{G}_{I}\phi \mid \phi_1 \bm{U}_{I} \phi_2 
\end{align} 
where $\mu : \mathbb{R}^n \rightarrow \mathbb{B}$ is the predicate whose boolean truth value is determined by the sign of a function $h : \mathbb{R}^n \rightarrow \mathbb{R}$, $\top$, $\neg$, $\land$, and $\lor$ are Boolean \textit{true}, \textit{negation}, \textit{and}, and \textit{or} operators, respectively, and $\bm{F}_{I}$, $\bm{G}_{I}$, $\bm{U}_{I}$ are the temporal \textit{eventually}, \textit{always}, and \textit{until} operators \color{blue}defined on a time interval $I =[a, b] = \{ t \in \mathbb{Z}_{\geq 0}: a\leq t \leq b\}$ ($a$, $b\in \mathbb{Z}_{\geq 0}$). 

\color{blue}
We define the semantics of an STL formula $\phi$ with respect to the trajectory $\mathbf{x}:=x_{0:T}$ of the system \req{dynamics} at time $t$ as follows:  
\begin{align}
    &(\mathbf{x},t)\models \mu \Leftrightarrow h (x_t)>0\notag \\
    &(\mathbf{x},t)\models \neg \mu \Leftrightarrow \neg ((\mathbf{x},t)\models \mu) \notag\\
    &(\mathbf{x},t)\models \phi_1 \land \phi_2 \Leftrightarrow (\mathbf{x},t)\models \phi_1 \land (\mathbf{x},t)\models \phi_2 \notag \\
    &(\mathbf{x},t)\models \phi_1 \lor \phi_2 \Leftrightarrow (\mathbf{x},t)\models \phi_1 \lor (\mathbf{x},t)\models \phi_2 \notag \\
    & (\mathbf{x},t)\models \bm{F}_{I}\phi \Leftrightarrow \exists t_1 \in t+I, (\mathbf{x},t_1)\models \phi\notag \\
    & (\mathbf{x},t)\models \bm{G}_{I}\phi \Leftrightarrow \forall t_1 \in t+I, (\mathbf{x},t_1)\models \phi \notag \\
     & (\mathbf{x},t)\models \phi_1 \bm{U}_{I}\phi_2 \Leftrightarrow \exists t_1 \in t+I\  \mathrm{s.t.}\  (\mathbf{x},t_1)\models \phi_2 \notag \\
     & \qquad \qquad \land \forall t_2 \in [t,t_1], (\mathbf{x},t_2)\models \phi_1\notag,
\end{align}
where $t+ I = \{t+k \in \mathbb{Z}_{\geq 0}: k \in I \}$.
For simplicity, we denote $\mathbf{x} \models \phi$ to abbreviate $(\mathbf{x}, 0) \models \phi$. 
The above semantics is \textit{qualitative}, in the sense that it reveals only if the trajectory $\mathbf{x}$ either satisfies or violates $\phi$. 
\color{black}

The notion of \textit{{robustness}} of STL formulas provides \textit{{quantitative}} semantics, and it measures {how much} the trajectory satisfies the STL formula. The robustness is sound in the sense that positive robustness value implies satisfaction of STL formula and negative robustness implies violation of STL formula. The robustness score of the STL formula $\phi$ is defined with respect to a trajectory $\mathbf{x}$ and a time $t$, which we denote by $\rho^\phi (\mathbf{x}, t)$, and is recursively defined as follows: 
\textcolor{blue}{
\begin{align}
    & \rho^\mu(\mathbf{x},t)&& \color{blue}= h (x_t) \notag \\
    & \rho^{\neg \mu}(\mathbf{x},t) &&= -h (x_t)\notag \\ 
    & \rho^{\phi_1 \land \phi_2}(\mathbf{x},t) &&= \min (\rho^{\phi_1} (\mathbf{x},t), \rho^{\phi_2} (\mathbf{x},t))\notag \\
    & \rho^{\phi_1 \lor \phi_2}(\mathbf{x},t) &&= \max (\rho^{\phi_1} (\mathbf{x},t), \rho^{\phi_2} (\mathbf{x},t))\notag \\
    & \rho^{\bm{F}_{I}\phi}(\mathbf{x},t) &&= \max_{t_1\in t+I}\rho^\phi(\mathbf{x},t)\notag \\
    & \rho^{\bm{G}_{I}\phi}(\mathbf{x},t) &&= \min_{t_1\in t+I}\rho^\phi(\mathbf{x},t)\notag \\
     & \rho^{\phi_1 \bm{U}_{I}\phi_2}(\mathbf{x},t) &&= \max_{t_1\in t+I}\Bigl(\min (\rho^{\phi_2}(\mathbf{x},t_1), \notag  \\
    & && \qquad \quad \min_{t_2 \in [t, t_1]}\rho^{\phi_1}(\mathbf{x},t_2))\Bigr)\notag
\end{align}
}
Note that the trajectory length $T$ should be selected large enough to determine the robustness score (see, e.g., \cite{MPC1}). 
For simplicity, we denote $\rho^{\phi}(\mathbf{x})$ to abbreviate $\rho^{\phi}(\mathbf{x}, 0)$. 

\color{black}

\section{Problem Statement}\label{problemstatement}
Let us now formulate a problem that we seek to solve throughout the paper. First, let $\Phi = \{\phi_1, \phi_2, \ldots, \phi_M\}$ denote a set of $M$ \textit{candidate} STL specifications. We assume that the STL specification can be freely chosen from the $M$ candidate specifications by the user before the control execution. Once the STL specification is chosen by the user, say $\phi_i \in \Phi$, the system \req{dynamics} is then controlled aiming to satisfy $\phi_i$. We also assume for simplicity that the chosen STL specification is fixed during control execution (as detailed below). 

In this paper, we aim to learn a feedback controller that maximizes the robustness of the STL specification. 
Note that the satisfaction and the robustness of an STL specification are defined over the trajectory of the system \req{dynamics}, and that the STL specification is freely chosen from $\Phi$. Hence, we should design a control policy that depends not only on the past and present states, but also on the STL specification, i.e., we need to obtain a control policy of the form $u_t = \pi (x_{0:t}, \phi_i; W)$, where $x_{0:t}=x_0,x_1,\dots,x_t$ is the trajectory of the system \req{dynamics}, $\phi_i$ is the STL specification that is chosen from $\Phi$, and $W$ denotes a set of parameters to be learned to characterize the control policy $\pi$. 
More formally, the problem considered in this paper is defined as follows: 
\begin{problem}\label{Problem1}
Given the system (\ref{dynamics}), horizon length $T$, probability distribution of initial states $p\ :\ \mathcal{X}_0\rightarrow \mathbb{R}$, and the candidate STL specifications $\Phi = \{\phi_1,\phi_2,\dots, \phi_M\}$, find a set of parameters $W$ that is the solution to the following problem:
\begin{align}
\mathop{\mathrm {maximize}}_{{W}} \ &\frac{1}{M} \sum_{i=1}^M \mathbb{E}_{p(x^i _0)} \left [ \rho^{\phi_i} \left(x_{0:T}^i\right)\right] \label{prob1 cost}\\
\mathrm{s.t.} \ \ & {x}_{t+1}^i = {f}\left({x}_t^i,\pi \left(x_{0:t}^i,\phi_i;{W}\right)\right),\label{prob1 const} \\
&t=0,1,\dots, T-1, \ i=0,1,\dots, M. \notag
\end{align}
where $\pi(\cdot,\cdot;W)$ denotes a control policy parameterized by $W$ and $x_{0:t}^i = x^i _0, x^i _1, \ldots, x^i _t$ is the trajectory of \req{dynamics} along with the policy $\pi(\cdot,\phi_i;W)$. 
\qedwhite \\  
\end{problem}

\begin{figure}[tb]
 \begin{center}
  \includegraphics[width=0.8\hsize]{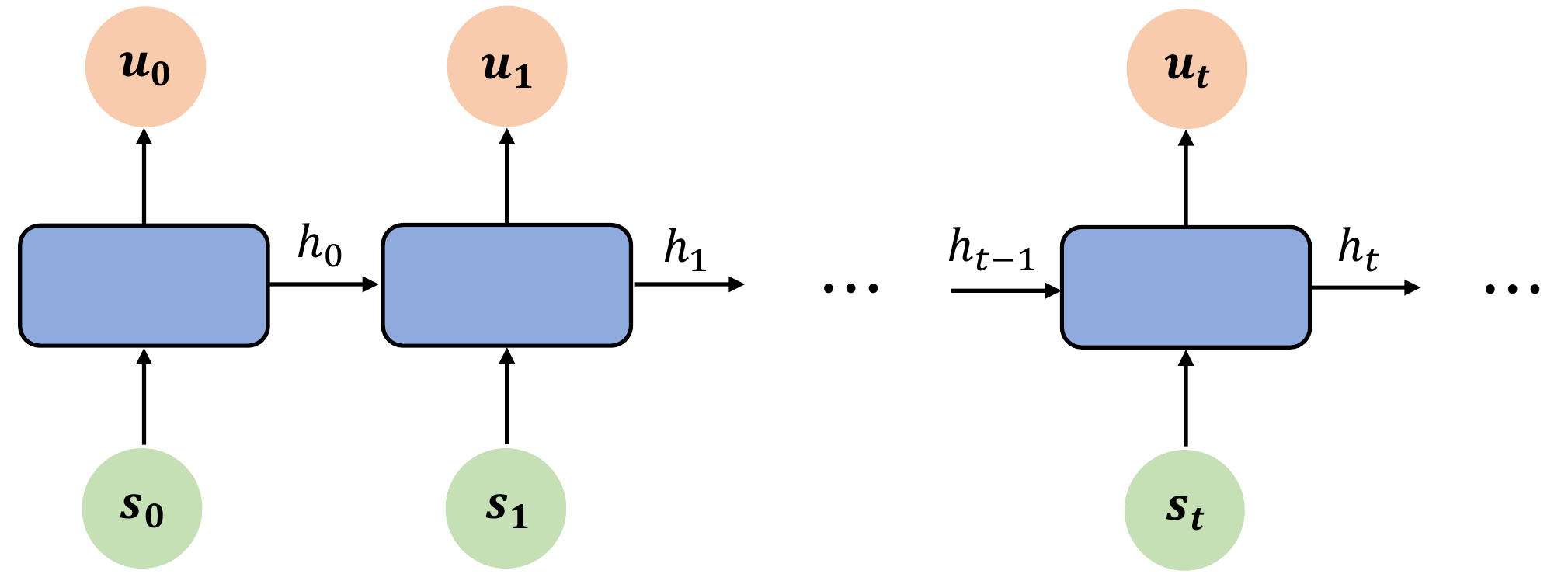}
 \end{center}
 \caption{The structure of RNN.}
 \label{RNN}
\end{figure}

In Problem \ref{Problem1}, we look for a set of the parameters of the control policy $W$ that maximizes the sum of the expectation of the robustness score with respect to the distribution of initial states over all $M$ candidate STL specifications. 
Since the control policy depends on the past and present states, or it is \textit{history dependent}, in this paper we use a recurrent neural network (RNN) to learn the control policy $\pi$ (see, \cite{RNN1,RNN2}). 
RNN is a type of a neural network that has a feedback architecture. A basic structure of the RNN is illustrated in \rfig{RNN}. 
As shown in \rfig{RNN}, RNN 
keeps processing the sequential information via internal hidden state $h$. The update rule of the hidden state and derivation of the output at time $t$ are as follows: 
\begin{align}
    h_t = g_{W_1} (h_{t-1}, s_t),\ u_t= l_{W_2} (h_{t}), \label{hteq} 
\end{align}
where $g_{W_1}$ and $l_{W_2}$ are the functions parameterized by weights $W_1$ and $W_2$, respectively, and $s_t$ denotes the input that is fed to the RNN. Hence, the set of the RNN parameters is given by $\{W_1, W_2\}$. 
As discussed in \cite{NN}, we can restrict the output of the RNN (i.e., the control input $u_t$) within the lower bound $u_{\mathrm{min}}$ and the upper bound $u_{\mathrm{max}}$ by defining each element $i$ $(i=1,\dots, m)$ of the function $l_{W_2}$ as follows: 
\begin{align}
    u_{t,i} = u_{\mathrm{min},i} + \frac{u_{\mathrm{max},i}-u_{\mathrm{min},i}}{2}\left(\mathrm{tanh}\left([W_2h_t]_i\right)+1\right), \label{uteq}
\end{align}
where the subscript $i$ denote $i$-th element of the vector.
Using \req{uteq}, we can generate control inputs satisfying $u_t \in \mathcal{U}$. 
A detailed definition of $s_t$ in \req{hteq} as well as a concrete procedure to learn $W_1, W_2$ are elaborated in \rsec{proposed}. 

Indeed, solving \rpro{Problem1} is not trivial and challenging in the following sense. Since the control policy depends on an STL specification $\phi_i$, we should consider how to {transform} each STL specification $\phi_i \in \Phi$ into a corresponding \textit{vector}, so that it can be fed to the RNN as the inputs together with the state trajectory $x_{1:t}$. Intuitively, it may be desirable that we can provide similar inputs to the RNN if the two STL specifications are close to each other in terms of their control policies. 
For example, consider a simple case with $x \in \mathbb{R}$ and the STL candidate specifications 
\begin{align}
&\phi_1 = \bm{F}_{[0, 10]} (0 \leq x \leq 1),\ \phi_2 = \bm{F}_{[0, 11]} (0 \leq x \leq 1), \\ 
&\phi_3 = \bm{F}_{[0, 10]} (10 \leq x \leq 11), \\
&\phi_4 = \bm{F}_{[0, 10]} (10 \leq x \leq 11) \lor \bm{F}_{[0, 10]} (12 \leq x \leq 13). 
\end{align}
Intuitively, control policies to satisfy $\phi_1$ and $\phi_2$ may be almost the same (as we aim to control the system to the same region with almost the same time intervals), while the control policies to satisfy $\phi_1$ and $\phi_3$ may be quite different (as we aim to control the system to different regions). 
In addition, if we could find a control policy to satisfy $\phi_3$, this control policy also leads to the satisfaction of $\phi_4$. 
Hence, it is convenient that we could provide a certain \textit{mapping}, where $\phi_1$ and $\phi_2$ are mapped onto the vector points that are close to each other but far from those corresponding to $\phi_3$ and $\phi_4$. In addition, $\phi_3$ and $\phi_4$ are mapped onto the same vector points if a control policy to satisfy $\phi_3$ could be found. 

Motivated by the above intuition, in this paper we propose a novel scheme for constructing a \textit{{vector representation}} of the STL specifications, which is referred to as \textit{STL2vec}. 
The proposed approach is inspired by the notion of Word2vec \cite{skip-gram}, a vector representation of words that has been proposed in machine learning literature and widely utilized for natural language processing. A concrete procedure of the proposed approach is elaborated in \rsec{proposed}. 

\section{Summary of Word2vec (skip-gram)} \label{word2vecsec}
\begin{figure}[tb]
 \begin{center}
  \includegraphics[width=0.8\hsize]{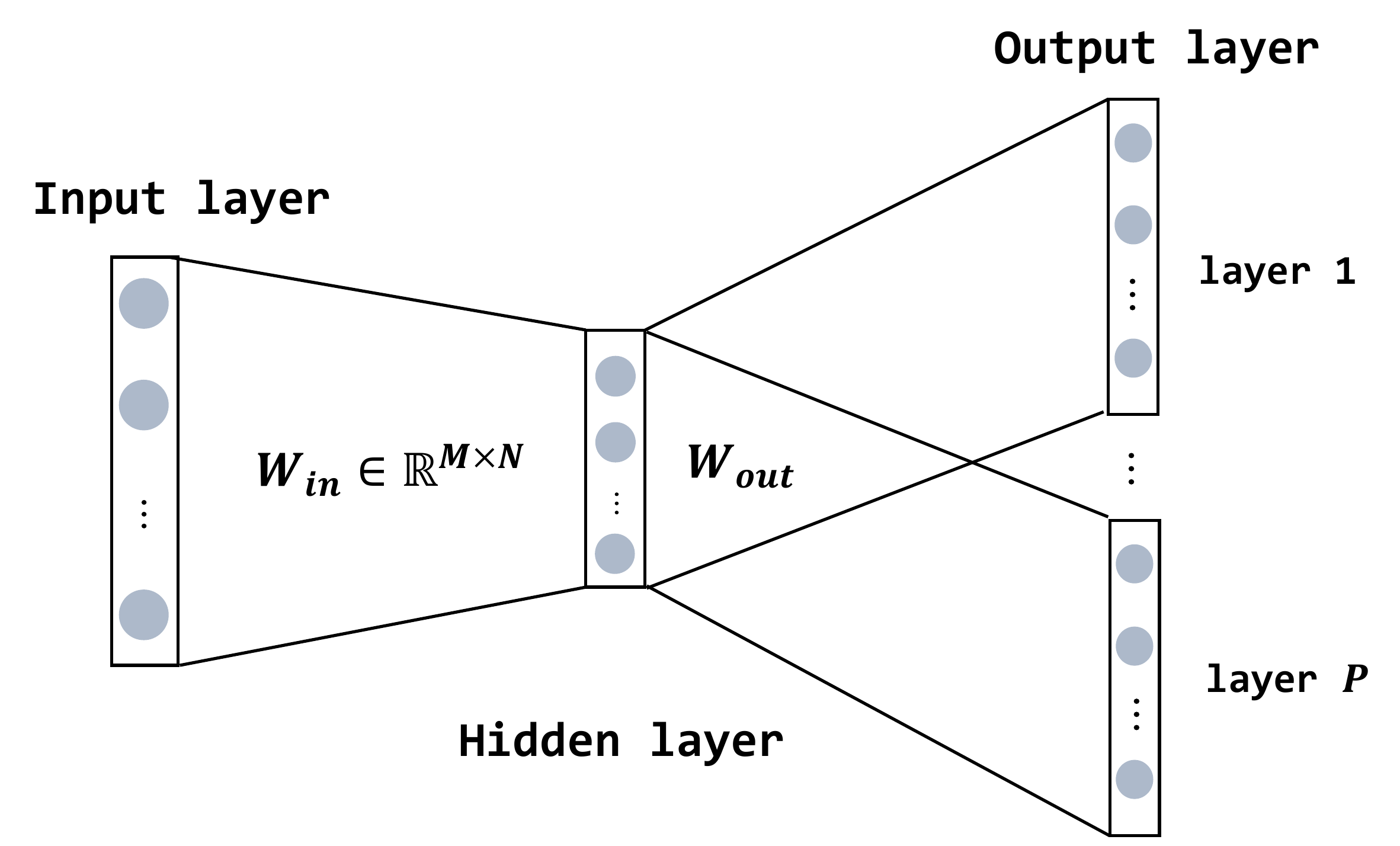}
 \end{center}
 \caption{Skip-gram model.}
 \label{Fig:skip-gram}
\end{figure}

Before providing the proposed approach, we summarize the basic concept of Word2vec. As we will see in the next section, the concept of Word2vec is a key ingredient to introduce STL2vec and how to train for it. 
The main objective of Word2vec is to group the vectors of similar or related words (for example, the words "man" and "boy" are mapped onto similar points in the vector space). As such, we can let the computer perform mathematical operations on words to detect their similarities. 
The mapping from each word to the vector is represented via an NN, i.e., the input for the NN is a word, and its output is a corresponding vector. One of the common techniques to train the NN is the \textit{skip-gram}\cite{skip-gram}. The structure of the skip-gram is shown in \rfig{Fig:skip-gram}. As shown in \rfig{Fig:skip-gram}, the skip-gram has a shallow three-layer NN. The input to the NN is an $M$-dimensional one-hot vector, i.e., only one single element is 1 and the other elements are all zero, where $M$ indicates a total number of words in the corpus (typically with $M > 10^6$). Here, each word is assigned by a unique one-hot vector (for example, "man" is labeled by $[1, 0, 0, \ldots, 0]^\mathsf{T}$, and "boy" is labeled by $[0, 1, 0, \ldots, 0]^\mathsf{T}$, \textit{etc}.), so that each word in the corpus is readable to the NN. 
The outputs of the NN are the collection of $P$ ($M$-dimensional) vectors whose each element represents the probability of the corresponding word in the corpus (training data for the output layer will be explained later in this section), where $P$ is a user-defined parameter. The mapping from the input layer to the hidden layer is given by a matrix $W_{\mathrm{in}}\in \mathbb{R}^{M\times N}$, where $N$ is a user-defined parameter that indicates the dimension of the word vector. 
The mapping from the hidden layer to each output layer is given by another matrix $W_{\mathrm{out}}\in \mathbb{R}^{N\times M}$. Here, we note that the matrix  $W_{\mathrm{out}}$ is the same for all the output layers. 
The weight parameters $W_{\mathrm{in}}, W_{\mathrm{out}}$ are trained from a large number of sentences, such as those in the news report (see, e.g. \cite{skip-gram}). For example, suppose $P=2$ and we have a sentence: 
"I want to eat an orange every morning". Then, we regard the training input as a certain word in this sentence and the training output as the set of $P=2$ words around it. For example, the word "eat" is regarded as the training input, and the set of $P=2$ words around it, i.e., \{to,\ an\} is regarded as the training outputs. 
Generating the training dataset as above is motivated by the insight that the meaning of a word is established by the surrounding words. 

Using the dataset constructed above, we train the skip-gram model. First, we pass the one-hot encoded input data to the input layer and perform the forward computation. As an output layer and loss function, we use softmax and cross-entropy loss, respectively. 
Then, we update the weights $W_{\mathrm{in}}$ and $W_{\mathrm{out}}$ using the gradients obtained by back-propagation computation (for details, see \cite{skip-gram}). After the training, the vector representation of each word in the corpus is given through the projection of the weight matrix $W_{\mathrm{in}}$ (hence, we drop the matrix $W_{\mathrm{out}}$ in the skip-gram model). That is, for each word $w$ in the corpus, the corresponding vector representation is given by \color{blue} $z_w=W_{\mathrm{in}} e_w \in \mathbb{R}^N$, where $e_w \in \mathbb{R}^M$ denotes the one-hot vector of the word $w$. \color{black}

\section{Proposed Scheme}\label{proposed}
\begin{figure}[tbp]
  \centering
  \subfigure[Step~(i): train STL2vec, whose input is the STL specification $\phi_i \in \Phi$ and the output is the $N$-dimensional vector.]{
    \centering
    \includegraphics[width = 0.6\hsize]{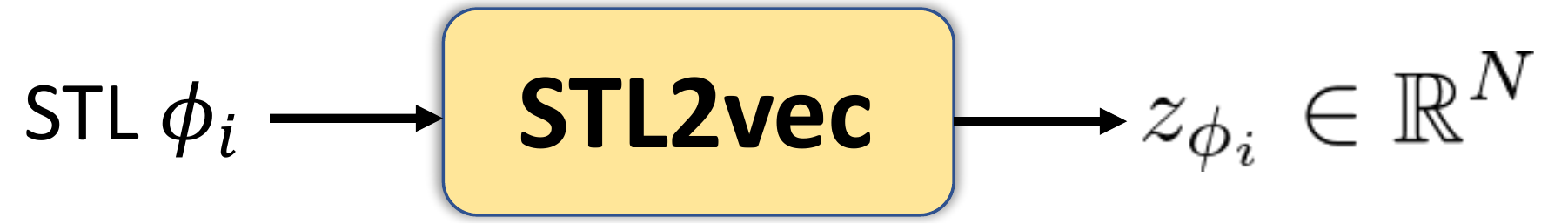}
    \label{step1}
    }\vspace{0cm}
  \subfigure[Step~(ii): train RNN, whose inputs are the state trajectory $x_{0:t}$ and the vector representation of the STL specification $z_{\phi_i}$, and the output is the control input $u_t$.]{
    \centering
    \includegraphics[width = 0.7\hsize]{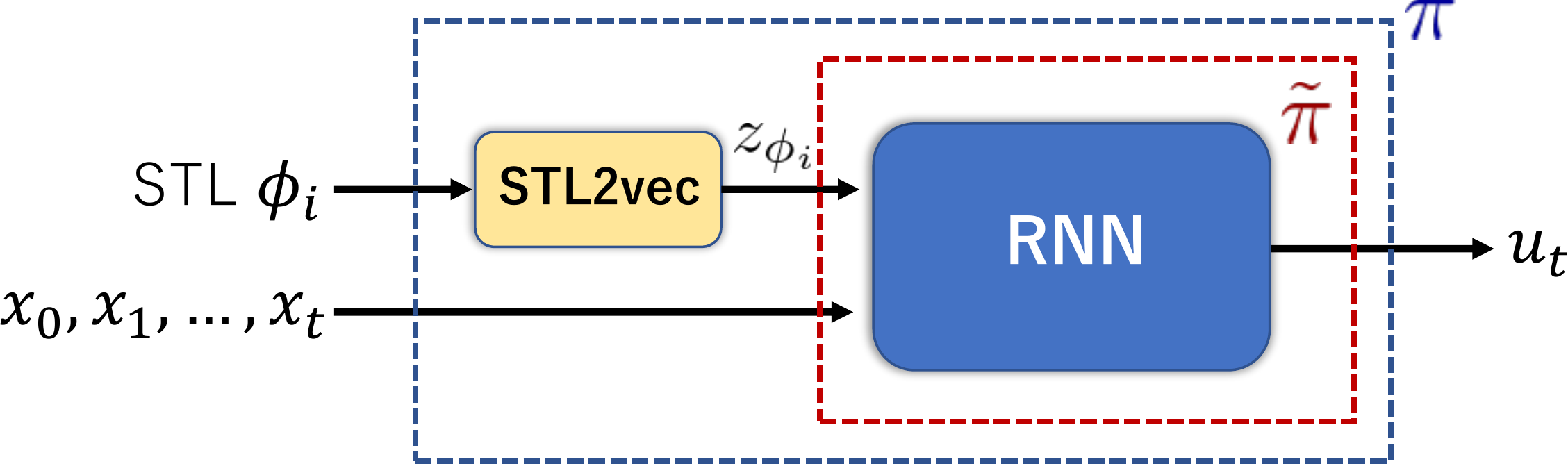}
    \label{step2}
  }
  \caption{Overview of the proposed approach.}
  \label{proposedapproachillust}
\end{figure}

In this section, we describe the solution approach to \rpro{Problem1}. 
The proposed approach consists of the two steps: (i) train a vector representation of the STL specifications, namely STL2vec, whose input is the STL specification $\phi_i \in \Phi$ and output is the $N$-dimensional vector; (ii) train the RNN, whose inputs are the state trajectory $x_{0:t}$ and the vector representation of the STL specification $z_{\phi_i}$, and the output is the control input $u_t$. The illustration of the two steps are shown in \rfig{proposedapproachillust}. Since STL2vec will be trained based on the skip-gram, the parameter to be learned in Step~(i) is $W_{\mathrm{in}}$ (recall in \rsec{word2vecsec} that we neglect $W_{\mathrm{out}}$). Also, recall that the parameters to be learned for RNN in Step~(ii) is $\{W_1, W_2\}$ (see \rsec{problemstatement}). Hence, the overall parameters to be learned for the control policy $\pi$ is $W = \{W_{\mathrm{in}}, W_1, W_2\}$. 

The proposed approach is beneficial in the following sense. By constructing STL2vec, we can obtain a vector representation that exhibits similarities between the STL specifications. This allows us to accelerate both efficiency and performance of the RNN controller in contrast to some naive approaches, \color{blue} such as integer or one-hot encoding i.e. simply assigning arbitrary numbers or one-hot vectors to the STL specifications (e.g., $\phi_1$ is assigned by 1, $\phi_2$ is assigned by 2, and so on) and use these numbers as the inputs to the RNN; for details, see an experimental result in \rsec{case study}. Moreover, the proposed approach is more beneficial than the approach of learning RNN controllers one by one for each STL formula in terms of computational memory; for details, see \rsec{remark:memory} and \rsec{case study}. 
\color{black}
The concrete procedures of the above two steps are given in the following subsections. 

\subsection{Training STL2vec}\label{STL2vecsec}
In this subsection, we provide a detailed procedure of Step~(i) (train STL2vec). To train STL2vec, we use the skip-gram model as explained in \rsec{word2vecsec}. 
First, we encode all the STL candidate specifications by the $M$-dimensional one-hot vectors, i.e. each STL specification $\phi_i \in \Phi$ is encoded by an $M$-dimensional vector whose $i$-th element is 1, and all other elements are 0 ($\phi_1$ is encoded as $[1, 0, \ldots, 0]^\mathsf{T}$ and $\phi_2$ is encoded as $[0, 1, 0, \ldots, 0]^\mathsf{T}$, and so on), so that they are readable to the NN. For simplicity, the one-hot vector of $\phi_i \in \Phi$ is denoted by $e_{\phi_i} \in \{0,1\}^M$. 

A key question regarding the construction of STL2vec is how to learn the weight parameters $W_{\mathrm{in}}, W_{\mathrm{out}}$ for the skip gram model, or in other words, how to generate the training dataset to learn $W_{\mathrm{in}}, W_{\mathrm{out}}$. 
In this paper, we learn these parameters such that similar vector representations can be obtained if the two STL specifications are {close} to each other in terms of their control policies (for a detailed intuition for this, see \rsec{problemstatement}). 
To this end, we generate training dataset by comparing \textit{robustness scores} among the STL specifications, aiming at measuring their closeness. 
More specifically, for each $\phi_i \in \Phi$, we randomly select an initial state $x_0 \in \mathcal{X}_0$ (following the probability distribution $p : \mathcal{X}_0 \rightarrow \mathbb{R}$), and solve the following maximization problem:
\begin{align}
&\mathop{\mathrm {maximize}}_{u_0, \ldots, u_{T-1}}\ \rho^{\phi_i} \left(x_{0:T}\right) \notag \\
&\mathrm{s.t.} \ \ {x}_{t+1} = {f}\left({x}_t,u_t\right),\ u_t \in \mathcal{U},\  t=0,1,\dots, T-1. \label{opt2}
\end{align}
Let $u_0^{*}, u_1^{*},\dots, u_{T-1}^{*} \in \mathcal{U}$ denote the optimal control inputs as the solution to \req{opt2}, and 
$x_{0:T}^{*}=x_0^{*}, x_1^{*},\dots, x_T^{*}$ the corresponding trajectory of the system \req{dynamics}. Thus, the corresponding (maximized) robustness score is $\rho^{\phi_i} \left(x^* _{0:T}\right)$. Then, we compute the robustness scores of the obtained trajectory $x_{0:T}^{*}$ with respect to \textit{all the {other}} STL specifications, i.e.,  
\begin{align}
\rho^{\phi_j}(x_{0:T}^{*}), \ \mathrm{for\ all}\ \phi_j \in \Phi\ \mathrm{with}\ i\neq j. \label{otherrobustness}
\end{align}
Then, we sort the robustness scores of \req{otherrobustness} in order of their closeness to $\rho^{\phi_i} \left(x^* _{0:T}\right)$ (i.e., sort by evaluating $|\rho^{\phi_i} \left(x^* _{0:T}\right) - \rho^{\phi_j} \left(x^* _{0:T}\right)|$, $i \neq j$). 
We denote the ordered specifications by $\phi^i _{k_1},\phi^i _{k_2},\dots, \phi^i _{k_{M-1}}$, where $\phi^i _{k_j} \in \Phi$ denotes the specification which has $j$-th closest robustness value to $\rho^{\phi_i}(x_{0:T}^{*})$.

\begin{algorithm}[t]\label{alg:constructionSTL2vec}
{\small
\SetKwInOut{Input}{Input}
\SetKwInOut{Output}{Output}
\Input{$\Phi = \{\phi_1, \ldots, \phi_M\}$ (candidate specifications); $x_0$ (initial state); $P$ (number of output layers); $N$ (dimension of vector representation); $N_{\mathrm{ite}}$ (number of iterations)}
\Output{$W_{\mathrm{in}}, W_{\mathrm{out}}$ (weight parameters for the skip-gram)} 
$\mathcal{D} \leftarrow \varnothing$; \\
\For{ each $\phi_i \in \Phi$ }{

\For{$\ell = 1 : N_{\mathrm{ite}}$ }{
Select $x_0 \in \mathcal{X}_0$ according to the probability distribution $p : \mathcal{X}_0 \rightarrow \mathbb{R}$; \\
Given $\phi_i \in \Phi$, $x_0 \in \mathcal{X}_0$, solve \req{opt2} to obtain the optimal trajectory $x^* _{0:T} = x^* _0, \ldots, x^* _T$; \\ 
Compute the robustness scores of $x^* _{0:T}$ with respect to the other STL specifications \req{otherrobustness}; \\ 

Sort \req{otherrobustness} in order of their closeness to $\rho^{\phi_i} \left(x^* _{0:T}\right)$, and pick up the first $P$ STL specifications: $\phi_{k_1}^{i},\phi_{k_2}^{i},\dots, \phi_{k_P}^{i}$; \\ 
Let $\mathcal{D}_{\mathrm{temp}}$ be given by \req{datatemp}. Then, update the dataset as $\mathcal{D} \leftarrow \mathcal{D} \cup \mathcal{D}_{\mathrm{temp}}$;}
}
Based on the training data set $\mathcal{D}$ and the cross entropy loss, train $W_{\mathrm{in}}, W_{\mathrm{out}}$ via back propagation. 
    \caption{Learning $W_{\mathrm{in}}, W_{\mathrm{out}}$ in the skip-gram.} 
    }
\end{algorithm}
Then, we pick up the first $P$ STL specifications, i.e., $\phi_{k_1}^{i},\phi_{k_2}^{i},\dots, \phi_{k_P}^{i}$ (recall that $P$ is the number of output layers in the skip-gram model), and set the input and output training data as $\left\{{\phi_i},\left ( {\phi^i_{k_1}}, {\phi^i_{k_2}}, \dots, {\phi^i_{k_P}} \right)\right \}$, where $\phi_i$ represents the input data and $(\phi^i_{k_1}, \phi^i_{k_2}, \dots, \phi^i_{k_P})$ represents the output data. This data is then encoded by the one-hot vectors, regarded as the training data to learn the skip-gram:
\begin{align}\label{datatemp}
\left\{e_{\phi_i},\left ( e_{\phi^i_{k_1}}, e_{\phi^i_{k_2}}, \dots, e_{\phi^i_{k_P}} \right)\right \}. 
\end{align}
The data in \req{datatemp} is then added to the dataset. 
\color{blue}
For each $\phi_i \in \Phi$, we repeat the above process $N_{\mathrm{ite}}$ times so as to obtain a collection of the training data.
\color{black}
The proposed approach presented above is summarized in \ralg{alg:constructionSTL2vec}.
Then, the weight parameters 
$W_{\mathrm{in}}, W_{\mathrm{out}}$ are learned by minimizing the cross entropy loss via back-propagation. 

\color{black}
\begin{remark}\label{remark:smooth}
\color{blue}{Note that, due to the definition of the robustness given in Section~\ref{pre}, the robustness can be nested with max/min functions and thus the optimization problem (\ref{opt2}) can be non-differentiable. 
To deal with such a problem, we adopt a \textit{smooth approximation} of the min/max operators by the log-sum-exp as follows: $\max (a_1,\dots,a_m)  \approx \frac{1}{\beta}\mathrm{ln} \sum_{i=1}^m \exp (\beta \alpha_i)$ and $\min (a_1,\dots,a_m) \approx \frac{1}{\beta}\mathrm{ln} \sum_{i=1}^m \exp (-\beta \alpha_i)$,
where $\beta >0$ is the scaling parameter. It is known that the approximation error goes to $0$ as $\beta \rightarrow \infty$ (see, e.g., \cite{smooth1}). 
This approximation allows the problem (\ref{opt2}) to be solved based on gradient-based algorithms.} \qedwhite 
\end{remark}

\begin{remark}
\color{blue}
Since the optimization problem (\ref{opt2}) is non-linear, finding the global optimal solution of the problem (\ref{opt2}) is difficult and some local optima that leads to undesirable control performance can be obtained. To the best of our knowledge, there are no general, systematic methodologies to obtain the global optimum of (\ref{opt2}). However, we believe that there are some heuristic methods that potentially improve the solution. For example, in the case study of Section~VI, we have confirmed that we could increase the possibility to obtain the reasonable solution (the solution with the positive robustness) by appropriately setting the parameter $\beta$ in the smooth approximation. 
Furthermore, we have also confirmed that we could deal with the problem by solving (\ref{opt2}) several times with the initial states randomly sampled from the vicinity of $x_0$, i.e., once $x_0$ is chosen, (\ref{opt2}) is solved with the initial state $\tilde{x}_0$ newly sampled satisfying $\|\tilde{x}_0-x_0\| \leq \epsilon$, where $\epsilon$ is a given small positive constant, which could increase the possibility to obtain positive robustness. 
\qedwhite 
\end{remark}
\color{black}



\color{black}
Finally, similarly to Word2vec, the vector representation of the STL specifications is given through the projection of $W_{\mathrm{in}}$, i.e., the vector representation of $\phi_i \in \Phi$ is given by $z_{\phi_i} = W_{\mathrm{in}} e_{\phi_i}$. 
\begin{example}\label{ex1}
Consider $M=5$ candidate STL specifications $\Phi = \{\phi_1, \phi_2,\dots, \phi_5\}$ and $P = 2$. 
Suppose that we consider $\phi_1$ and solve the corresponding optimization problem \req{opt2} (i.e., $i = 1$) to obtain the optimal trajectory $x_{0:T}^{*}$. 
Suppose that the robustness scores of $x_{0:T}^{*}$ with respect to the STL specifications are obtained as  $\rho^{\phi_1}(x_{0:T}^{*})=0.3$, $\rho^{\phi_2}(x_{0:T}^{*})=0.2$, $\rho^{\phi_3}(x_{0:T}^{*})=-0.5$, $\rho^{\phi_4}(x_{0:T}^{*})=0.1$, and $\rho^{\phi_5}(x_{0:T}^{*})=0.25$. 
Then, we select the top $P=2$ specifications among $\{\phi_2, \phi_3, \phi_4, \phi_5 \}$ whose robustness scores are the closest to the one of $\phi_1$. 
In this case, we select $\phi_5$ and $\phi_2$, since the robustness scores of $\phi_5$ and $\phi_2$ have the most and the second closest robustness values to $\rho^{\phi_1}(x_{0:T}^{*})$, respectively. Thus, the resulting data is $\{(\phi_1),(\phi_5,\phi_2)\}$ and the encoded training data to be added in the dataset is $\left \{[1,0,0,0,0],\ \left([0, 0, 0, 0, 1], [0,1,0,0,0]\right)\right \}$. 
 \qedwhite 
\end{example}

\smallskip 
\begin{remark}
If there exist two or more STL specifications that have the same robustness values in the output data, we add all the combinations to the training data set.
For example, in Example~1, if the robustness values are obtained as  $\rho^{\phi_1}(x_{0:T}^{*})=0.3$, $\rho^{\phi_2}(x_{0:T}^{*})=0.2$, $\rho^{\phi_3}(x_{0:T}^{*})=-0.5$, $\rho^{\phi_4}(x_{0:T}^{*})=0.2$, and $\rho^{\phi_5}(x_{0:T}^{*})=0.1$, then we add the training data as $\left\{ (\phi_1),(\phi_2,\phi_5) \right\}$ and $\left\{ (\phi_1),(\phi_4,\phi_5)\right\}$. \qedwhite 
\end{remark}

\subsection{Training RNN}\label{trainRNNcontrollersec}
Here, we describe a detailed procedure of Step~(ii). Recall that in Step~(ii), we aim to train RNN, whose inputs are the trajectory $x_{0:t}$ and the vector representation of $\phi_i$ (i.e., $z_{\phi_i}$), and the output is the control input $u_t$. 
Further, remember that the concrete unfolded structure of the RNN is shown in \rfig{RNN}, and the update rules are given by \req{hteq}--\req{uteq}. 
Here, the input variable $s_t$ is given by the concatenation of the state $x_t$ and the vector representation of the chosen STL specification, i.e., $s_t = [x_t^\mathsf{T}, z_{\phi_i}^\mathsf{T}]^\mathsf{T}$, where $z_{\phi_i} \in \mathbb{R}^N$ is the vector representation of the chosen STL specification $\phi_i$. 
Now, let $\tilde{\pi}(x_{0:t}, z_{\phi_i} ;W_1, W_2)$ denote the control policy (or a mapping) for the RNN, where $W_1, W_2$ are the RNN parameters to be learned (for the illustration, see \rfig{step2}). 
Since the parameters for STL2vec $W_{\mathrm{in}}$ is fixed, we here focus on learning the RNN parameters $W_1, W_2$ to characterize $\tilde{\pi}$. 
\color{blue}

To numerically solve the Problem \ref{Problem1}, we keep repeating the following procedure until a certain number of epochs $N_{\mathrm{epo}}$ is reached. 
First, in each epoch, we construct \textit{mini-batch} of the specifications by randomly rearranging the order of the candidate specifications and splitting them up according to the prespecified batch size $N_b$, i.e., we construct the batches $\mathcal{B}_1 = \{ \phi_{1}^1,\ldots, \phi_{N_b}^1\}$, $\mathcal{B}_2=\{ \phi_{1}^2,\ldots, \phi_{N_b}^2\}$, $\dots$, $\mathcal{B}_K=\{ \phi_{1}^K,\ldots, \phi_{N_b}^K\}$, where $K=\frac{M}{N_b}$ (assuming that $M$ can be divided by $N_b$) \footnote{If $M$ is not divisible by $N_b$, we construct minibatch with $K=\lfloor\frac{M}{N_b}\rfloor$ ( $\lfloor \cdot \rfloor$ denotes a floor function) and append the remaining specifications in the last batch. For example, if $\Phi = \{\phi_1, \phi_2, \ldots, \phi_5\}$, and $N_b = 2$, we construct minibatch e.g., $\mathcal{B}_1 = \{\phi_1, \phi_3\}$, $\mathcal{B}_2 = \{\phi_2, \phi_4, \phi_5\}$.}.
Then, we iterate the following procedures for all the batches $\mathcal{B}_p\ (p=1,\ldots,K)$ to update the parameters $W_1$ and $W_2$.

\subsubsection{Forward Computation}
For all the pairs of the initial state $x_0^j$ $(j=1, \ldots, L)$ which are randomly sampled from the initial region $\mathcal{X}_0$ and the vectors of the specifications $z_{\phi_{i}^p}$ $(i=1, \ldots, N_b)$ in a batch $\mathcal{B}_p$,
we generate the trajectories $x_{0:T}^{i,j}$ by iteratively applying  ${x}_{t+1}^{i,j}={f}\left({x}_t^{i,j}, \tilde{\pi} \left(x^{i,j}_{0:t}, z_{\phi_{i}^p};{W_1, W_2}\right)\right)$ for fixed $W_1$ and $W_2$    
from the initial state $x_0^{i,j}=x_0^j$.
Then, by using all the $N_b L$ trajectories obtained above, we compute the following loss:
\begin{align}
    -\frac{1}{{N_b}L} \sum_{i=1}^{{N_b}} \sum_{j=1}^L \rho^{\phi_{i}^p} \left(x_{0:T}^{i,j}\right),
\end{align}
which is an approximation of the expectation (\ref{prob1 cost}) in Problem~\ref{Problem1} with the specifications in a batch (note that we take the negative of the robustness to define a loss).
\color{black}
\subsubsection{Backward Computation}
After the forward computation, we compute the gradients for all the parameters via backpropagation through time (BPTT) \cite{LSTM}. This computation can be easily implemented by combining the auto-differentiation tools designed for NNs like PyTorch \cite{pytorch} and STLCG \cite{stlcg} which is a newly developed python toolbox for computing the STL robustness using computation graph. 
\subsubsection{Weight Update}
Lastly, we update all the parameters in the RNN controller by using the weights obtained above. In this paper, we use Adam optimizer \cite{adam}.

\color{blue}
\subsection{Some comparisons between the proposed approach and the approach of learning the RNNs one by one}\label{remark:memory}
As one of the alternative methods, one could consider learning a set of RNN controllers independently \textit{one by one} for all the STL specifications. 
This one-by-one approach might eventually provide higher control performance than the proposed method, since it tries to learn a set of RNN controllers independently for all the candidate STL specifications (on the other hand, the proposed method learns the controller by using only one RNN model). However, the proposed approach has the potential to require much less computational memory than the one-by-one approach due to the following reason. 
In the proposed method, the total number of the parameters to be learned is given by $N_{\mathrm{in}}+N_{R1}+N_{R2}$, where $N_{\mathrm{in}}$, $N_{R1}$, and $N_{R2}$ are the number of elements in $W_{\mathrm{in}}$, $W_1$, and $W_2$.
For instance, when we use the Long-Short-Term-Memory (LSTM) as the RNN model, $N_{R1}$ and $N_{R2}$ are given by $N_{R1}=4N_{h}N_{\mathrm{lstm}}(n+N)$ and $N_{R2}=4N_{h}N_{\mathrm{lstm}}m$, 
where $N_h$ is the dimension of the hidden state of RNN and $N_{\mathrm{lstm}}$ is the number of LSTM layers. The number of parameters for STL embedding $N_{\mathrm{in}}$ is given by $N_{\mathrm{in}}=MN$.
On the other hand, the required number of parameters when we train the controller one-by-one for each candidate specifications is given by $M(N^{'}_{R1}+N^{'}_{R2})$, where $N^{'}_{R1}=4N_{\mathrm{lstm}}nN_h$ and $N^{'}_{R2}=N_{R2}=N_{\mathrm{lstm}}mN_{\mathrm{h}}$ are the number of parameters for each RNN model when we train STL specifications one-by-one. 
From the above discussion, the number of the parameters for the STL embedding $N_{\mathrm{in}}$ and the RNN models in the one-by-one case are \textit{both} proportional to $M$, and the corresponding coefficients are given by $N$ and $N^{'}_{R1}+N^{'}_{R2}$, respectively. In other words, the number of the required parameters increases with respect to the number of the STL specifications in both of the two approaches. 
However, it is noted that we typically have $N \ll N^{'}_{R1}+N^{'}_{R2}$\footnote{\textcolor{blue}{It is argued that $N \ll N^{'}_{R1}+N^{'}_{R2}$ indeed holds for many control problems with STL tasks based on the fact that the previous works regarding the control synthesis with RNN (e.g., in \cite{RNN2}) use the structure of 2 LSTM layers with 32 dimensional hidden states, which leads to $N^{'}_{R1}+N^{'}_{R2}=1280$, even for simple 2-D path planning problem. 
On the other hand, in the literature of word2vec \cite{skip-gram}, vector size $N$ for word embedding is typically set from 50 to 200 for very large corpus like $780\times 10^3$ (in our case study of Section~VI, we have set $N=20$ that is smaller than these typical values, since the total number of the candidate STL specifications considered in the case study is $369 \ll 780\times 10^3$).}}, i.e., the size of the vector representation of STL is selected much smaller than the number of the RNN parameters, and thus the number of the required parameters by the proposed approach is smaller than the one-by-one approach. For example, in the case study of Section~VI, it is shown that the total number of parameters to learn the controller by the proposed approach is 24 times smaller than the one-by-one approach; for details, see Section~\ref{case study}. 
\color{black}

\section{case study}\label{case study}
\begin{figure}[tb]
 \begin{center}
  \includegraphics[width=0.6\hsize]{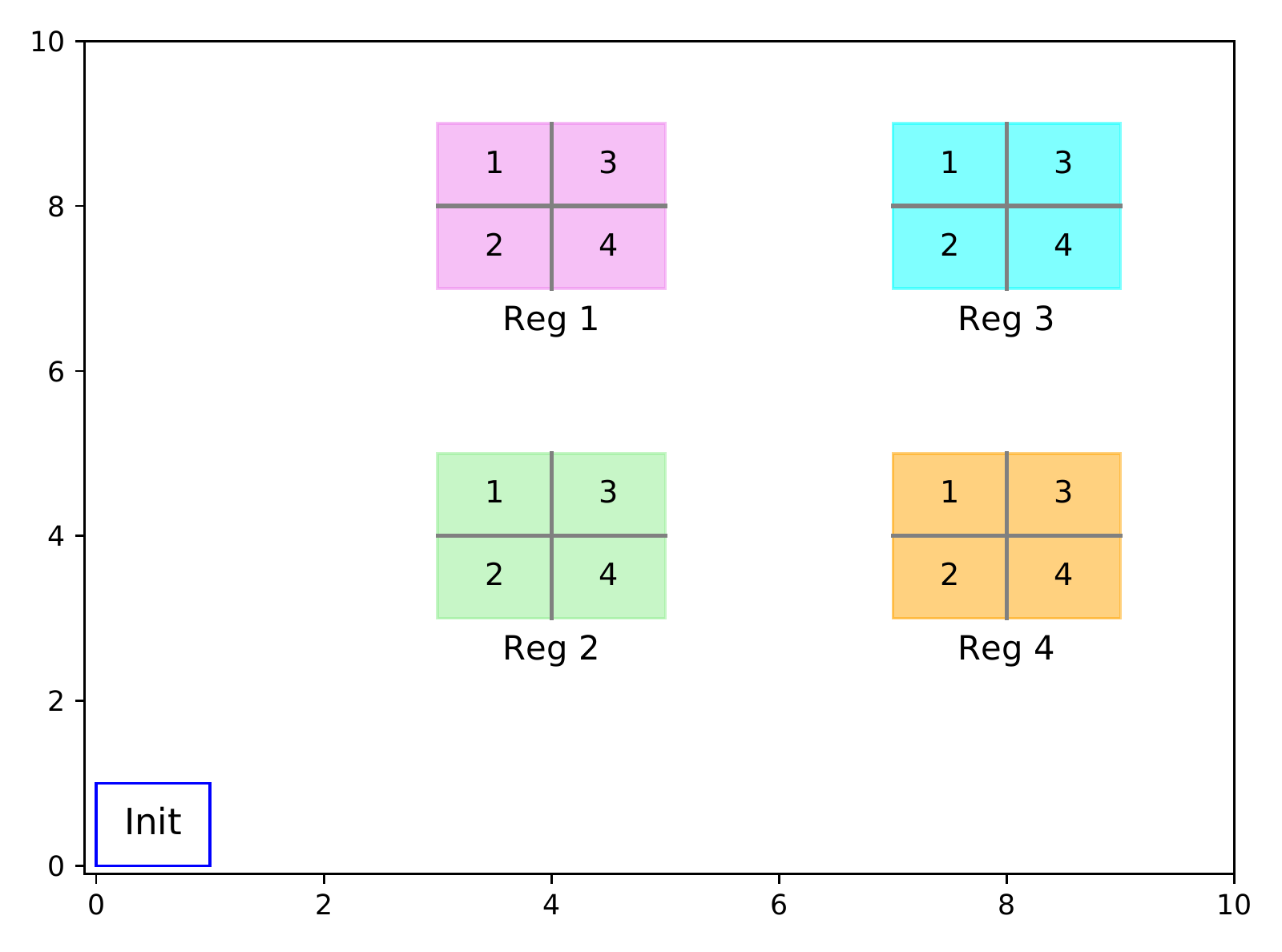}
 \end{center}
 \caption{Regions and sub-regions in a 2D space}
 \label{field}
\end{figure}

We show the efficacy of the proposed method through the example of a path planning problem of a vehicle in 2D space. We used PyTorch package \cite{pytorch} for the implementation of RNN, \color{blue}IPOPT in CasADi \cite{casadi} to solve optimization (\ref{opt2}), \color{black} and STLCG \cite{stlcg} for the computation regarding STL robustness.  
\color{blue}
We used Windows~10 with a 2.80 GHz Core i7 CPU and 32 GB of RAM for all the experiments discussed in this section. 
\color{black}
Throughout this section, we consider the following nonlinear, discrete-time nonholonomic system: $q_{x,t+1}=q_{x,t} + v_t\sin{\theta_t}$, $q_{y,t+1}=q_{y,t} + v_t\cos{\theta_t}$, $\theta_{t+1}=\theta_t+\omega_t$, 
where $q_{x,t}$, $q_{y,t}$ represent position of the vehicle, $\theta_t$ is the heading angle, $v_t$ and $\omega_t$ are velocity and angular velocity, respectively. The state $x_t$ and control input $u_t$ are defined as $x_t=[q_{x,t},q_{y,t},\theta_t]^\top$ and $u_t = [v_t, \omega_t]^\top$, respectively. 
As illustrated in Figure \ref{field}, we consider 4 regions in a 2D space: $\mathrm{Reg} \ 1=[3,5]\times [7,9]$, $\mathrm{Reg}\ 2=[3,5]\times [3,5]$, $\mathrm{Reg} \ 3=[7,9]\times [3,5]$, $\mathrm{Reg}\ 4=[7,9]\times [7,9]$, and the set of initial states $\mathcal{X}_0 = [0, 0.7]\times [0, 0.7]$ (blue-edged color region). In addition, we consider 4 sub-regions in each region, which are labeled by indices as shown in Figure \ref{field}. In the following, we denote $j$-th sub-region $(j=1,2,3,4)$ in region $i$ $(i=1,2,3,4)$ as $\mathrm{Reg}\ (i,j)$. 

\color{blue}
The STL candidate specifications are shown in Table~\ref{table:STLs}. Here, the specifications are considered for all $i, i_1, i_2, i_3 \in \{1, \ldots, 4\}$ ($i_1 > i_2$, $i_3\neq 1$, $i_1\neq i_3$) and $j, j_1, j_2, j_3 \in \{1, \ldots, 4\}$. 
The total number of the STL candidate specifications is given by $369$. 
\color{black}
\subsection{Training STL2vec}\label{Section STL2vec}
Now we train STL2vec based on the proposed approach presented in \rsec{STL2vecsec}. 
\begin{table}[tb]
  \caption{STL candidate specifications.}
  \label{table:STLs}
  \centering
  \begin{tabular}{l}
    \hline
      Specifications \\
    \hline 
    $(a)$     $\bm{F}_{[0,20]}\ \mathrm{Reg}\ (i,j)$\\
    $(b)$     $\bm{F}_{[0,20]}\ \mathrm{Reg}\ (i_1,j_1) \lor \bm{F}_{[0,20]}\ \mathrm{Reg}\ (i_2,j_2)$\\
    $(c)$     $\bm{F}_{[0,10]}\ \mathrm{Reg}\ (i_1,j_1) \land \bm{F}_{[11,20]}\ \mathrm{Reg}\ (i_3,j_2)$\\
    $(d)$     $\bm{F}_{[0,15]}\bm{G}_{[0,5]}\ \mathrm{Reg}\ (i,j)$\\
    $(e)$     $\bm{F}_{[0,15]}\bm{G}_{[0,5]}\ \mathrm{Reg}\ (i_1,j_1) \lor \bm{F}_{[0,15]}\bm{G}_{[0,5]}\ \mathrm{Reg}\ (i_2,j_2)$\\
   \textcolor{red}{$(f)$    $\bm{F}_{[0,20]}\ \mathrm{Reg}\ (4,4) {\land} \left(\lnot \mathrm{Reg}\ (4,4)\ \bm{U}_{[0,20]}\ \mathrm{Reg}\ (2,3)\right)$}\\
    \hline
  \end{tabular}
\end{table}
\begin{table}[tb]
  \caption{The time required to obtain STL2vec (in sec)}
  \label{run-time}
  \centering
  \begin{tabular}{lc}
    \hline
     Procedure & Run-time\\
     \hline
     Dataset generation (STL2vec)& 1452  \\
     Training STL2vec& 87  \\
    \hline
  \end{tabular}
\end{table}
The parameters for the skip-gram are \color{black}$N=20$ and \color{black}$P=2$.
$N_{\mathrm{ite}}$ in Algorithm~\ref{alg:constructionSTL2vec} and the number of epochs for the training of skip-gram model are set to 1 and 100, respectively.
After the training, we evaluate similarities between all the 2 different vector representations $z_{\phi_i},z_{\phi_j}$ ($i\neq j$) by the {cosine similarity}, which is defined as $(z_{\phi_i}^\mathsf{T} z_{\phi_j})/(\|z_{\phi_i}\|\|z_{\phi_j}\|)$ and it takes the maximum value of $1$ if $z_{\phi_i}$ has the same orientation as $z_{\phi_j}$. 
The time required to train STL2vec is summarized in Table \ref{run-time}.
In Table~\ref{table:Similarity}, we illustrate STL specifications which have the largest to fourth-largest cosine similarity values for some example specifications.


We summarize some characteristics that we have observed in the obtained embeddings in the followings: 
(I) Each specification in (a) of Table~\ref{table:STLs} is typically embedded close to the corresponding specification in (d). Moreover, each specification in (b) and (e) is embedded close to 
either $\bm{F}_{[0,20]}\ \mathrm{Reg}\ (i_1,j_1)$ (and $\bm{F}_{[0,15]}\bm{G}_{[0,5]}\mathrm{Reg}\ (i_1,j_1)$) or $\bm{F}_{[0,20]}\ \mathrm{Reg}\ (i_2,j_2)$ (and $\bm{F}_{[0,15]}\bm{G}_{[0,5]}\mathrm{Reg}\ (i_2,j_2)$). Typically, (b) and (e) are embedded close to the specification regarding the region which is closer to init region.
For example, as we can partially see in Ex 1 of Table \ref{table:Similarity}, we have observed that almost all the specifications $\bm{F}_{[0,20]}\ \mathrm{Reg}\ (2,2) \lor \bm{F}_{[0,20]}\ \mathrm{Reg}\ (i,j)$ and $\bm{F}_{[0,15]}\ \bm{G}_{[0,5]}\ \mathrm{Reg}\ (2,2) \lor \bm{F}_{[0,15]}\ \bm{G}_{[0,5]}\ \mathrm{Reg}\ (i,j)$ $(i\in \{1,3,4\},\ j \in \{1,2,3,4\})$ are embedded close to the specification $\bm{F}_{[0,20]}\ \mathrm{Reg}\ (2,2)$.  
This is intuitive and desirable result when we train the controller since all of these specifications are satisfied by entering whether $\mathrm{Reg}\ (i_1,j_1)$ or $\mathrm{Reg}\ (i_2,j_2)$ and stay there more than 5 steps, which is possible actions for all of regions in this example.
(II) Each specification in (c) of Table~\ref{table:STLs} is basically embedded close to the specifications which are satisfied by similar trajectory.
Specifically, the properties same as the following examples are observed for all the specifications in (c): (i) specifications $\bm{F}_{[0,10]}\ \mathrm{Reg}\ (1,3) \land \bm{F}_{[11,20]}\ \mathrm{Reg}\ (3,j)$ ($j\in \{1,2,3,4\}$) are embedded close to each other (see Ex 2 of Table \ref{table:Similarity}); (ii) as we can see in Ex 3 of Table \ref{table:Similarity}, the specifications $ \bm{F}_{[0,10]}\ \mathrm{Reg}\ (2,2) \land \bm{F}_{[11,20]}\ \mathrm{Reg}\ (3,1)$ and $\bm{F}_{[\textcolor{red}{0},20]}\ \mathrm{Reg}\ (3,1)$ are mapped onto similar vectors (these specifications are satisfied with the same trajectory since $\mathrm{Reg}(2,2)$ is on the way from the starting point to $\mathrm{Reg}(3,1)$).
(III) The specification (f) of Table~\ref{table:STLs} is embedded close to $ \bm{F}_{[0,10]}\ \mathrm{Reg}\ (2,3) \land \bm{F}_{[11,20]}\ \mathrm{Reg}\ (4,4)$ as we can see from EX 4 in Table \ref{table:Similarity}. 
This is also desirable result result since the specifications (f) require to firstly reach $\mathrm{Reg}\ (2,3)$ and then reach $\mathrm{Reg}\ (4,4)$ after that.
{
\begin{table*}[tb]
  \caption{Cosine similarities of vector representations of the STL candidate specifications.}
  \label{table:Similarity}
  \centering
  \begin{tabular}{clc||clc}
    \hline \hline
     Ex 1&$\bm{F}_{[0,20]}\ \mathrm{Reg}\ (2,2)$& $\mathrm{sim}$ & Ex 2&$\bm{F}_{0,10]}\ \mathrm{Reg}\ (1,3) \land \bm{F}_{[11,20]}\ \mathrm{Reg}\ (3,1)$& $\mathrm{sim}$\\
    \hline \hline
    $1$ & $ \bm{F}_{[0,20]}\ \mathrm{Reg}\ (3,3) \lor \bm{F}_{[0,20]}\ \mathrm{Reg}\ (2,2)$ & 0.99
    & $1$ & $\bm{F}_{[0,10]}\ \mathrm{Reg}\ (1,3) \land \bm{F}_{[11,20]}\ \mathrm{Reg}\ (3,4)$ & 0.99 \\
   
    $2$ & $ \bm{F}_{[0,20]}\ \mathrm{Reg}\ (2,2) \lor \bm{F}_{[0,20]}\ \mathrm{Reg}\ (1,1)$& 0.99 &$2$ & $\bm{F}_{[0,10]}\ \mathrm{Reg}\ (1,3) \land \bm{F}_{[11,20]}\ \mathrm{Reg}\ (3,3)$ & 0.99\\
    
    $3$ & $ \bm{F}_{[0,15]}\bm{G}_{[0,5]}\ \mathrm{Reg}\ (4,4) \lor \bm{F}_{[0,15]}\bm{G}_{[0,5]}\ \mathrm{Reg}\ (2,2)$ & 0.99 & $3$ & $\bm{F}_{\textcolor{red}{[0,10]}}\ \mathrm{Reg}\ (1,3) \land \bm{F}_{\textcolor{red}{[11,20]}}\ \mathrm{Reg}\ (3,2)$ & 0.95\\
    
    $4$ & $ \bm{F}_{[0,15]}\bm{G}_{[0,5]}\ \mathrm{Reg}\ (2,2)$ & 0.99 & $4$ & $ \bm{F}_{[0,15]}\bm{G}_{[0,5]}\ \mathrm{Reg}\ (3,3) \lor \bm{F}_{[0,15]}\bm{G}_{[0,5]}\ \mathrm{Reg}\ (1,3)$ & 0.76\\

    \hline \hline
     Ex 3&$ \bm{F}_{[0,10]}\ \mathrm{Reg}\ (2,2) \land \bm{F}_{[11,20]}\ \mathrm{Reg}\ (3,1)$ & $\mathrm{sim}$ & Ex 4&\textcolor{red}{$\bm{F}_{[0,20]}\ \mathrm{Reg}\ (4,4) {\land} \left(\lnot \mathrm{Reg}\ (4,4)\ \bm{U}_{[0,20]}\ \mathrm{Reg}\ (2,3)\right)$}& $\mathrm{sim}$\\
    \hline \hline
    $1$ & $ \bm{F}_{[0,20]}\ \mathrm{Reg}\ (3,1)$ & 0.96 & $1$ & $ \bm{F}_{[0,10]}\ \mathrm{Reg}\ (2,3) \land \bm{F}_{[11,20]}\ \mathrm{Reg}\ (4,4)$ &0.81 \\
    
    $2$ & $ \bm{F}_{[0,20]}\ \mathrm{Reg}\ (3,1)\lor \bm{F}_{[0,20]}\ \mathrm{Reg}\ (1,1)$ & 0.95 & $2$ & $ \bm{F}_{[0,10]}\ \mathrm{Reg}\ (4,4) \land \bm{F}_{[11,20]}\ \mathrm{Reg}\ (3,3)$ & 0.73\\
    
    $3$ & $ \bm{F}_{[0,20]}\ \mathrm{Reg}\ (4,4)\lor \bm{F}_{[0,20]}\ \mathrm{Reg}\ (3,1)$ & 0.95 & $3$ & $ \bm{F}_{[0,10]}\ \mathrm{Reg}\ (4,4) \land \bm{F}_{[11,20]}\ \mathrm{Reg}\ (3,1)$ & 0.71\\
    
    $4$ & $ \bm{F}_{[0,15]}\bm{G}_{[0,5]}\ \mathrm{Reg}\ (4,4)\textcolor{black}{\lor} \bm{F}_{[0,15]}\bm{G}_{[0,5]}\ \mathrm{Reg}\ (3,1)$ & 0.93 & $4$ & $ \bm{F}_{[0,10]}\ \mathrm{Reg}\ (4,4) \land \bm{F}_{[11,20]}\ \mathrm{Reg}\ (3,4)$ & 0.71\\

    \hline \hline
    
  \end{tabular}
\end{table*}}

\subsection{Training RNN}
Next, we evaluate the control performance of the proposed method.
In this experiment, we train the parameters of RNN model for the specifications (a), (b), (c), (f) and $\bm{F}_{[0,15]}\bm{G}_{[0,5]}\ \mathrm{Reg}\ (1,1) \lor \bm{F}_{[0,15]}\bm{G}_{[0,5]}\ \mathrm{Reg}\ (3,2)$ in (e) in Table \ref{table:STLs}.
As for the specifications in (c), we only consider the specifications with $(i_1,i_3) = (1,3), (2,1), (2,3), (2,4), (4,3)$. Thus the total number of specifications is 194. We retrain the STL embeddings (20 dimension) with these specifications and use them as the input to the RNN.
Same as the previous work \cite{RNN2}, all of the RNN models used in this example consists of 2 LSTM layers with 32-dimensional hidden states. 
\begin{table}[tb]
  \caption{Some alternative approaches.}
  \label{table:naive}
  \centering
  \begin{tabular}{l}
    \hline
     Approaches\\
     \hline
     (A1)   Integer encoding scheme\\
     (A2)  One-hot encoding scheme\\
     (A3) Training controllers one-by-one\\
    \hline
  \end{tabular}
\end{table}
As summarized in Table \ref{table:naive}, we consider the following 3 alternative approaches for comparison with the proposed method: (A1) Integer encoding scheme: we incrementally assign an integer number for each STL specification (for example, $\bm{F}_{[0,20]}\ \textcolor{red}{\mathrm{Reg}}\ (1,1)$ is assigned by $1$, $\bm{F}_{[0,20]}\ \textcolor{red}{\mathrm{Reg}}\ (1,2)$ is assigned by $2$, \textit{etc}), and use it as the input to the RNN (instead of the vectors generated by STL2vec) (A2) One-hot encoding scheme: we generate and assign 194 dimensional one-hot vectors for all the 194 specifications (for example,  $\bm{F}_{[0,20]}\ \textcolor{red}{\mathrm{Reg}}\ (1,1)$ is assigned by $[1,0,\dots,0]$, $\bm{F}_{[0,20]}\ \textcolor{red}{\mathrm{Reg}}\ (1,2)$ is assigned by $[0,1,\dots,0]$, \textit{etc})  (A3) Training one-by-one: we ready 194 RNN models and train each specification one-by-one.
When we train the controllers for the approaches (A1) and (A2), we used the same training procedure discussed in \rsec{trainRNNcontrollersec} and used integer numbers or one-hot vectors instead of the vectors obtained by STL2vec. In both the proposed approach and (A1), (A2), we set the batch size and the number of initial states sampled for training in each iteration ($N_b$ and $L$ in \rsec{trainRNNcontrollersec})) to 8 and 3, respectively. 
In (A3), for each epoch we generate initial states $x^j_0$ ($j=1,\ldots,L=3$) randomly from $\mathcal{X}_0$ and update the RNN parameters  assigned for each STL specification $\phi_i \in \Phi$ ($i=1, \ldots,M$) one by one via forward/backward computation (similarly to the procedure of Section~V-B) with the following loss: $-\frac{1}{L}\sum_{j=1}^L\rho^{\phi_{i}}\left(x_{0:T}^{j}\right)$.
\begin{table}[tb]
  \caption{Run time needed to reach several average of robustness (in sec).}
  \label{result: time}
  \centering
  \begin{tabular}{lcccc}
    \hline
     approach \textbackslash average of robustness & 0.1 & 0.15 & 0.2 & 0.22\\
     \hline
     Proposed method& 3494 & 4054 & 6722 & 14668 \\
    Learning one-by-one& 5686 & 7663 & 10336 & 12168 \\
     One-hot encoding& 27238 & 33863 & -- & --\\
     Integer encoding& -- & -- & -- &--\\
     \hline
  \end{tabular}
\end{table}

\begin{figure}[tb]
 \begin{center}
  \includegraphics[width=0.7\hsize]{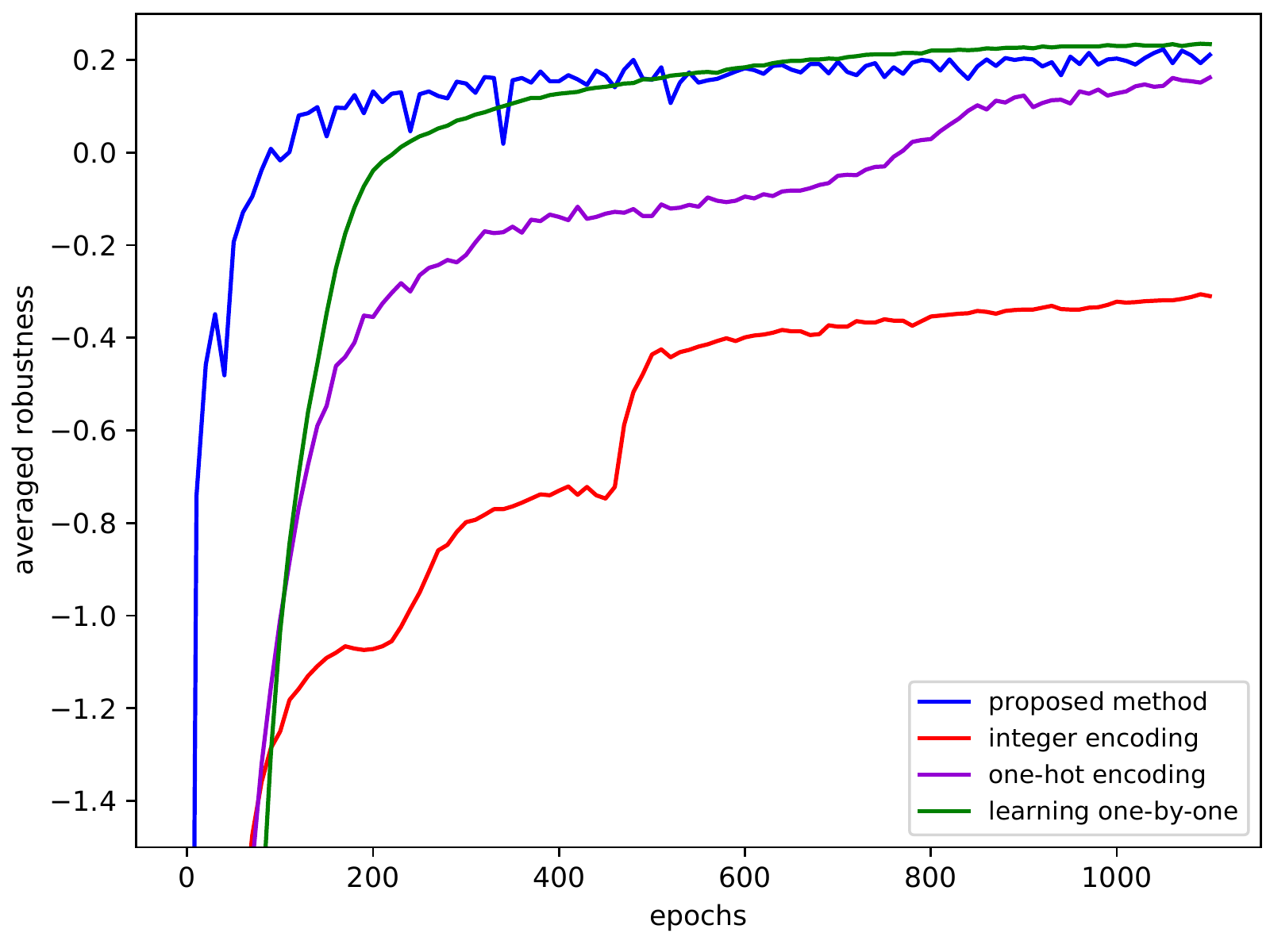}
 \end{center}
 \caption{The average of robustness values for each control scheme (proposed method and schemes (A1)-(A3)). The average has been taken by testing the controllers for every 10 epochs with 30 initial states randomly sampled from the initial region $\mathcal{X}_0$.}
 \label{result: robustness}
\end{figure}

\begin{figure}[tbp]
  \centering
  \subfigure[Trajectories for the proposed method.]{
    \centering
    \includegraphics[width = 0.85\hsize]{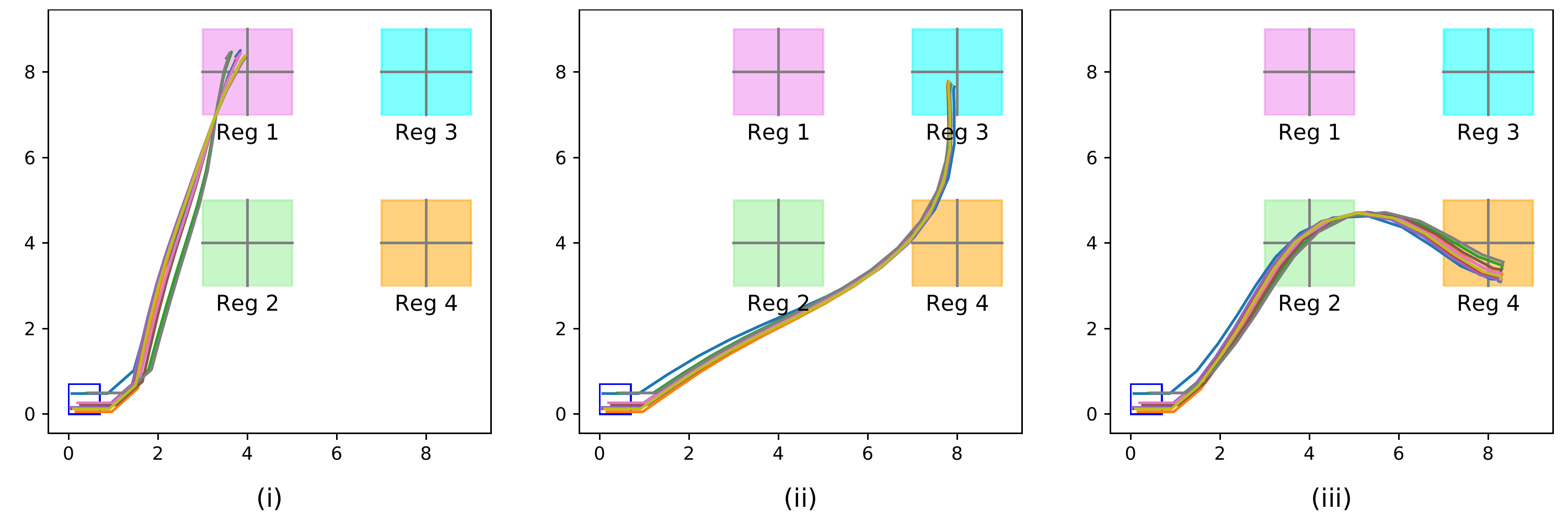}
    \label{exp proposed}
    }\vspace{0cm}
  \subfigure[Trajectories for the approach (A3)]{
    \centering
    \includegraphics[width = 0.85\hsize]{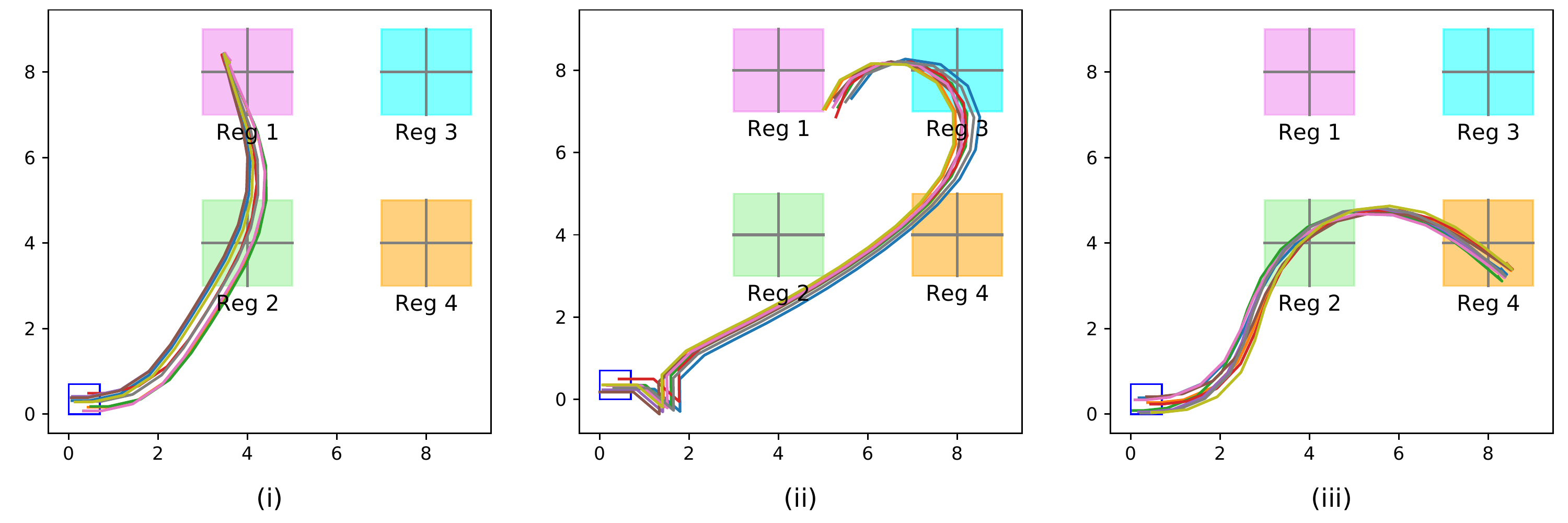}
    \label{exp naive}
  }
  \caption{The testing result for 10 initial states and the 3 specifications (i) $F_{[0,15]}G_{[0,5]}\textcolor{red}{\mathrm{Reg}}\ (1,1)\lor F_{[0,15]}G_{[0,5]}\textcolor{red}{\mathrm{Reg}}\ (3,2)$, (ii) $F_{[0,10]}\textcolor{red}{\mathrm{Reg}}\ (4,1)\land F_{[\textcolor{red}{11},20]}\textcolor{red}{\mathrm{Reg}}\ (3,2)$, (iii) $\bm{F}_{[0,20]}\ \mathrm{Reg}\ (4,4) \land \left(\lnot \mathrm{Reg}\ (4,4)\ \bm{U}_{[0,20]}\ \mathrm{Reg}\ (2,3)\right)$ are plotted. The results are shown for both (a) proposed scheme and (b) approach (A3).}
  \label{exp contol}
\end{figure}
Fig.~\ref{result: robustness} shows the average of robustness values for all the control schemes and Table~\ref{result: time} summarizes the actual time (in sec) required to reach the robustness values 0.1, 0.15, 0.2, 0.22 for all the proposed and alternative methods (the symbol "--" indicates that the corresponding average of the robustness value has not been achieved). 
Furthermore, the resulting few trajectories obtained by applying the RNN controller trained by the proposed method and approach (A3) are plotted in \rfig{exp contol} (both controllers were trained 1100 epochs and the trajectories are plotted for 10 initial states newly sampled from $\mathcal{X}_0$).
The robustness values are collected by testing the controllers once every 10 epochs with 30 initial states newly sampled from the initial region $\mathcal{X}_0$. 
The values in Fig. \ref{result: robustness} for the one-by-one scheme (A3) are plotted by taking the mean of the average of the robustness values for all the 194 RNN controllers at the same epochs. 
We further note here that the actual times required to run one epoch for all the approaches are different from each other. The averaged times required to run the proposed approach and the approaches (A1)-(A3) were 14.2, 14.0, 32.2, and 14.0 [s], respectively (note that the time for approach (A3) is the total sum of the times required to update the parameters of all the RNN models). The time required for the approach (A2) is larger than the other methods because of the large input dimension (the dimension of the one-hot vector is 194). On the other hand, we have confirmed that the time required to update the parameters is not so much increased for the proposed method when $N=20$ as shown above.
As we can see from Fig \ref{result: robustness} and Table~\ref{result: time}, the average of robustness value of the proposed method are improving faster at the beginning of the training procedure than the other methods in terms of both number of epochs and actual time. Especially, from Table~\ref{result: time}, we can confirm that the average of robustness value of the proposed method reaches 0.1, 0.15, and 0.2 faster than the other approaches. 
Moreover, as we mentioned in Subsection \ref{remark:memory}, we can largely save the memory consumption compared with the scheme (A3). In Table \ref{table: param}, we summarize the total number of the parameters required for each approach in this example. 

However, the average of robustness value of the proposed method is subtly overtaken by the approach (A3) around 500 epoch and the averaged robustness value of the approach (A3) reaches 0.22 faster than the proposed approach.
Within 1100 epochs, the maximum average of robustness values of the proposed method and approach (A3) were $0.233$ and $0.235$, respectively.
The reason why the average of robustness values of the proposed method is overtaken by that of the approach (A3) may be because of the effect of the other specifications, i.e., since only one RNN controller is trained for many specifications in the proposed method, the control performance for a specification may be affected by the other specifications depending on the obtained embedding. Such effect is observed in (b) of \rfig{exp contol}. The resulting trajectories are converged to the ones that are relatively far from the middle of $\mathrm{Reg}(3,2)$, which leads to the low robustness although the specification itself is satisfied. To remove such behavior, further investigation for obtaining more superior embedding will be one of our future works.




\
\begin{table}[tb]
  \caption{Total number of parameters in each method.}
  \label{table: param}
  \centering
  \begin{tabular}{lc}
    \hline
     & Number of parameters\\
     \hline
     Proposed method & 10280 \\
     approach (A1) & 1280 \\
     approach (A2) & 50432 \\
     approach (A3) & 248320 \\
    \hline
  \end{tabular}
\end{table}
\
\section{conclusion and Future work}\label{conclusion}
We proposed a method for mapping STL specifications onto the vector space (STL2vec) 
based on the word2vec technique. To obtain the STL embeddings that capture the similarities in specifications in terms of control policy, we have provided a method for constructing the dataset by solving the robustness maximization problem for all the candidate specifications. 
Then, we trained the RNN controller whose inputs are the state trajectory and a vector generated by STL2vec to deal with multiple STL specifications with one RNN model. 
The example shown in the simulation section shows efficacy of the proposed method in terms of memory consumption and the time required for the training.

\textcolor{blue}{In this paper, it is assumed that the chosen STL specification is fixed during control execution and not allowed to change during the execution. Hence, future work should involve investigating the case where the STL specification is changed during control execution so as to provide more flexibility of the RNN controller.} 

\section*{Acknowledgement}
This work is supported by JST CREST JPMJCR2012. 

\end{document}